\documentclass[twocolumn]{aastex701}

\usepackage{xcolor}
\usepackage{amsmath}
\usepackage{booktabs, tabularx, makecell,array,threeparttable}
\usepackage{hyperref}

\begin{document}
\title{Dust Properties of the Interstellar Object 3I/ATLAS Revealed by Optical and Near-Infrared Polarimetry}

\author[orcid=0009-0006-1028-1653]{Seungwon Choi}
\affil{Department of Physics and Astronomy, Seoul National 
University, 1 Gwanak-ro, Gwanak-gu, Seoul 08826, Korea}
\affil{SNU Astronomy Research Center, Department of Physics and Astronomy, Seoul National University, 1 Gwanak-ro, Gwanak-gu, Seoul 08826, Republic of Korea}
\email[show]{cygalbireo@snu.ac.kr}

\author[orcid=0000-0002-7332-2479]{Masateru Ishiguro}
\affiliation{Department of Physics and Astronomy, Seoul National 
University, 1 Gwanak-ro, Gwanak-gu, Seoul 08826, Korea}
\affiliation{SNU Astronomy Research Center, Department of Physics and Astronomy, Seoul National University, 1 Gwanak-ro, Gwanak-gu, Seoul 08826, Republic of Korea}
\email[show]{ishiguro@snu.ac.kr}

\author[orcid=0000-0002-2928-8306]{Jun Takahashi}
\affiliation{Nishi-Harima Astronomical Observatory, Center for Astronomy, University of Hyogo, 407–2 Nishigaichi, Sayo-cho, Hyogo 679–5313, Japan
}
\email{}

\author[]{Tomoki Saito}
\affiliation{Nishi-Harima Astronomical Observatory, Center for Astronomy, University of Hyogo, 407–2 Nishigaichi, Sayo-cho, Hyogo 679–5313, Japan
}
\email{}

\author[orcid=0000-0002-2618-1124]{Yoonsoo P. Bach}
\affiliation{Korea Astronomy and Space Science Institute (KASI), 776 Daedeok-daero, Yuseong-gu, Daejeon 34055, Republic of Korea}
\email{}

\author[orcid=0000-0002-8244-4603]{Bumhoo Lim}
\affiliation{Department of Physics and Astronomy, Seoul National 
University, 1 Gwanak-ro, Gwanak-gu, Seoul 08826, Korea}
\affiliation{SNU Astronomy Research Center, Department of Physics and Astronomy, Seoul National University, 1 Gwanak-ro, Gwanak-gu, Seoul
08826, Republic of Korea}
\email{}

\author[orcid=0000-0001-9067-7653]{Hiroyuki Naito}
\affiliation{Nayoro Observatory, 157-1 Nisshin, Nayoro, Hokkaido, 096-0066, Japan}
\email{}

\author[orcid=0000-0002-3291-4056]{Jooyeon Geem}
\affiliation{Asteroid Engineering Laboratory, Lule\r{a} University of Technology, Kiruna, Sweden}
\email{}

\author[orcid=0000-0002-0460-7550]{Sunho Jin}
\affiliation{Korea Astronomy and Space Science Institute (KASI), 776 Daedeok-daero, Yuseong-gu, Daejeon 34055, Republic of Korea}
\email{}

\author[]{Jinguk Seo}
\affiliation{Department of Physics and Astronomy, Seoul National 
University, 1 Gwanak-ro, Gwanak-gu, Seoul 08826, Korea}
\affiliation{SNU Astronomy Research Center, Department of Physics and Astronomy, Seoul National University, 1 Gwanak-ro, Gwanak-gu, Seoul
08826, Republic of Korea}
\email{}

\author[orcid=0009-0001-8488-1786]{Hyeonwoo Ju}
\affiliation{Department of Physics, Konkuk University, 120 Neungdong-ro, Gwangjin-gu, Seoul 05029, Korea}
\email{}

\author[orcid=0000-0001-6156-238X]{Hiroshi Akitaya}
\affiliation{Astronomy Research Center, Chiba Institute of Technology, 2-17-1 Tsudanuma, Narashino, Chiba 275-0016, Japan}
\affiliation{Planetary Exploration Research Center, Chiba Institute of Technology, 2-17-1 Tsudanuma, Narashino, Chiba 275-0016, Japan}
\affiliation{Hiroshima Astrophysical Science Center, Hiroshima University, 1-3-1 Kagamiyama, Higashi-Hiroshima, Hiroshima 739-8526, Japan}
\email{}

\author[orcid=0000-0001-6099-9539]{Koji S. Kawabata}
\affiliation{Hiroshima Astrophysical Science Center, Hiroshima University, 1-3-1 Kagamiyama, Higashi-Hiroshima, Hiroshima 739-8526, Japan}
\email{}

\author[orcid=0000-0001-5946-9960]{Mahito Sasada}
\affiliation{Institute of Integrated Research, Institute of Science Tokyo, 2-12-1 Ookayama, Meguro-ku, Tokyo 152-8550, Japan}
\email{}

\author[orcid=0009-0007-2559-4609]{Kazuya Doi}\affiliation{Graduate School of Science, Hokkaido University, Kita-ku, Sapporo, Hokkaido 060-0810, Japan}
\email{}

\author[orcid=0000-0003-0693-1618]{Hisayuki Kubota}
\affiliation{Department of Earth and Planetary Sciences, Faculty of Science, Hokkaido University, Kita-ku, Sapporo, Hokkaido 060-0810,
Japan}
\email{}

\author[orcid=0000-0002-7084-0860]{Seiko Takagi}
\affiliation{Department of Earth and Planetary Sciences, Faculty of Science, Hokkaido University, Kita-ku, Sapporo, Hokkaido 060-0810,
Japan}
\email{}

\author[orcid=0000-0002-3656-4081]{Makoto Watanabe}
\affiliation{Department of Physics, Okayama University of Science, 1-1 Ridai-cho, Kita-ku, Okayama, Okayama 700-0005, Japan}
\email{}

\author[orcid=0000-0003-1726-6158]{Tomohiko Sekiguchi}
\affiliation{Asahikawa Campus, Hokkaido University of Education, Hokumon 9, Asahikawa, Hokkaido 070-8621,Japan}
\email{}

\author[orcid=0000-0002-8537-6714]{Myungshin Im}
\affiliation{Department of Physics and Astronomy, Seoul National 
University, 1 Gwanak-ro, Gwanak-gu, Seoul 08826, Korea}
\affiliation{SNU Astronomy Research Center, Department of Physics and Astronomy, Seoul National University, 1 Gwanak-ro, Gwanak-gu, Seoul
08826, Republic of Korea}
\email{}

\newcommand{\Rc}{\ensuremath{R_\mathrm{C}}}
\newcommand{\Ic}{\ensuremath{I_\mathrm{C}}}
\newcommand{\Ks}{\ensuremath{K_\mathrm{s}}}

\begin{abstract}
We present independent polarimetric observations of the interstellar object 3I/ATLAS, including the first near-infrared polarimetric measurements. Using imaging polarimeters, we measured the degree of linear polarization from the visible \Rc-band (0.64~\micron) to the near-infrared \Ks-band (2.25~\micron), and investigated its dependence on solar phase angle (polarization phase curve; PPC) and wavelength (polarization color curve; PCC).
We confirm that the PPC of 3I/ATLAS differs significantly from those of typical Solar System comets, showing an unusually large polarization amplitude. This PPC shows no significant change in the \Rc\ band across perihelion passage, despite the perihelion lying within the water snow line.
This indicates that the unusual polarimetric behavior of 3I/ATLAS is unlikely to be driven by transient volatile activity, but instead reflects intrinsic optical properties of refractory dust particles. 
The PCC increases with wavelength over 0.6--1.2~\micron\ and peaks at 1.5--2.0~\micron, suggesting that the dominant scattering units are dust aggregates composed of submicron-sized monomers, broadly consistent with interstellar dust and solar-system cometary aggregates. Taken together, our results indicate that 3I/ATLAS preserves polarimetric properties characteristic of a primitive cometary planetesimal formed in another planetary system, with a refractory dust composition that differs from that typically observed among Solar System comets, despite sharing a similar size scale of the aggregate building blocks.
\end{abstract}


\section{Introduction} 

Polarimetric observations of small bodies in the Solar System, including comets, asteroids, trans-Neptunian objects, and interstellar objects (ISOs), provide a powerful diagnostic tool for probing the physical properties of scattering dust particles. In particular, the degree of linear polarization of scattered sunlight, together with its dependence on solar phase angle (the polarization phase curve, PPC) and wavelength (the polarization color curve, PCC), provides key constraints on particle size distributions, optical constants, porosity, and aggregate structure.

Historically, cometary polarimetry has been carried out primarily at visible wavelengths, with near-infrared measurements limited to a few exceptionally bright comets such as 1P/Halley, C/1995 O1 (Hale–Bopp), C/1975 V1 (West), and C/2023 A3 (Tsushinshan-ATLAS) \citep{1978PASJ...30..149O, 1987A&A...187..621B, Kikuchi1987, 1997EM&P...78..353H, 2025ApJ...983L..19L}. After early observational biases \citep{1996A&A...313..327L}, mainly due to contamination by molecular emission lines, were recognized and corrected, most Solar System comets (SSCs) were found to exhibit remarkably similar optical PPCs, largely independent of their dynamical origin in either the Oort cloud or the Kuiper belt. With the exception of a small number of highly evolved comets \citep{2019A&A...629A.121K,2004AJ....128.3061J}, this apparent uniformity suggests that the solid particles that formed cometary planetesimals were broadly homogeneous in the outer regions of the protoplanetary disk beyond the giant-planet formation zone, and that polarimetry is particularly sensitive to such primordial dust properties \citep{2025ApJ...983L..19L}.

The recent discovery of ISOs offers a unique opportunity to test whether this apparent uniformity of cometary dust properties is a universal outcome of planet formation or a peculiarity of the Solar System. Polarimetric observations of the second ISO, 2I/Borisov, revealed subtle but measurable deviations from the typical SSC polarization phase curves \citep{2021NatCo..12.1797B}. On 2025 July 1, the third known ISO, 3I/ATLAS, was discovered at a heliocentric distance of 4.49 au. Soon after its discovery \citep{2025MPEC....N...12D}, intensive polarimetric observations were initiated by \citet{2025ApJ...992L..29G}. They reported that the polarization degree of 3I/ATLAS differs significantly from that of SSCs before crossing the water snow line.
Furthermore, \citet{2025RNAAS...9..338Z} reported an apparent discontinuous change in its PPC across the water snow line, and suggested that the presence of volatile-rich material may play a key role in shaping the unusual polarimetric behavior of this object.

In this Letter, we present independent polarimetric observations of 3I/ATLAS. Notably, we report the first near-infrared polarimetric measurements of this ISO. Our data also include PPC measurements obtained near perihelion and during the outbound phase, which have not been reported previously. By combining these new observations with existing visible-wavelength data, we investigate the size and optical properties of the dust particles composing 3I/ATLAS, and discuss the implications for the diversity of planetesimal formation environments beyond the Solar System.

\section{Observations and Data Analysis}

The specifications of our observation and observational circumstances are summarized in Table~\ref{tab:observation_log}.
The optical imaging polarimetry was conducted with the Seoul National University (SNU) Quadruple Imaging Device for POLarimetry (SQUIDPOL) attached to the Cassegrain focus of the 0.6-m Ritchey--Chr\'etien telescope at the Pyeongchang Observatory of SNU \citep{2025epsc.conf..579G}, the Multi-Spectral Imager (MSI) attached to the 1.6-m Pirka telescope at the Nayoro Observatory of the Faculty of Science, Hokkaido University \citep{2012SPIE.8446E..2OW}, and the Hiroshima Optical and Near-InfraRed camera (HONIR) attached to the Cassegrain focus of the 1.5-m Kanata telescope at the Higashi-Hiroshima Observatory (HHO), Hiroshima University \citep{2014SPIE.9147E..4OA}. For MSI and HONIR, we utilized the polarimetric observation modes. 
For all optical observations, we chose the standard Johnson--Cousins \Rc\ and \Ic\ filters. Although HONIR is capable of optical and near-infrared observations, only the optical polarimetric data were analyzed in this study. We also conducted near-infrared imaging polarimetry using the polarimetric mode of the Nishiharima Infrared Camera (NIC) attached to the 2.0-m Nayuta telescope at the Nishi-Harima Astronomical Observatory (NHAO), University of Hyogo. NIC offers simultaneous $J$-, $H$- and \Ks-bands imaging, including imaging polarimetry (\citealt{Takahashi2019}).

All instruments used in this study are beam-splitter-type polarimeters, allowing the ordinary and extraordinary components to be sampled simultaneously. The fields of view and pixel scales of the instruments are as follows: SQUIDPOL has a field of view of $15.6 \times 10.7$~arcmin with a pixel scale of $0.45\arcsec$, HONIR in the imaging polarimetric mode consists of five slots each covering a field of view of $9.7 \times 0.75$~arcmin with a pixel scale of $0.29\arcsec$, MSI in the imaging polarimetric mode provides a field of view of $3.3 \times 0.7$~arcmin with a pixel scale of $0.39\arcsec$, and NIC in the imaging polarimetric mode offers a $24 \times 69$~arcsec field of view with a pixel scale of $0.16\arcsec$.


To compare the polarization degree at different wavelengths, some optical observations were coordinated with the near-infrared observations to be as simultaneous as possible. The datasets were selected considering weather and seeing conditions. All datasets were analyzed in a standard manner, consistent with previous works (\citealt{Takahashi2019}, \citealt{2022MNRAS.509.4128I}, \citealt{2024A&A...684A..81B}, \citealt{2025ApJ...983L..19L}). The data reduction procedure included preprocessing (bias, dark, and flat-field corrections), extraction of target signals, correction for instrumental polarization and polarization efficiency, derivation of the normalized Stokes parameters ($Q/I$ and $U/I$), and calculation of the polarization degree perpendicular to the scattering plane ($P_\mathrm{r}$).

\begin{deluxetable*}{ccccccccc}[h]
\tabletypesize{\small}
\tablewidth{\textwidth}
\setlength{\tablecolumns}{9}
\setlength{\tabcolsep}{6pt}
\tablecaption{Observation log.\label{tab:observation_log}}
\tablehead{
\colhead{UT Date} & \colhead{Instrument} & \colhead{Filter/Wavelength} & \colhead{Exptime [s]} & \colhead{N} & \colhead{$r_H$[au]} & \colhead{$\Delta$[au]} & \colhead{$\alpha$[\degr]}& \colhead{Airmass} \\
\colhead{(1)} & \colhead{(2)} & \colhead{(3)} & \colhead{(4)} & \colhead{(5)} & \colhead{(6)} & \colhead{(7)} & \colhead{(8)} & \colhead{(9)}
}
\startdata
\textbf{2025-11-13.85$^\dagger$} & NIC & $J$, $H$, \Ks & 60 & 5, 6, 5 & 1.47 & 2.13 & \textbf{24.16} & 2.58 \\
\textbf{2025-11-14.84$^\dagger$} & SQUIDPOL & \Rc, \Ic & 180, 300 & 4,4 & 1.48 & 2.11 & \textbf{24.87} & 4.19 \\
2025-11-15.85 & MSI & \Rc, \Ic & 60, 120 & 2, 4 & 1.50 & 2.10 & 25.55 & 2.44 \\
2025-11-18.85 & SQUIDPOL & \Rc & 180 & 3 & 1.55 & 2.06 & 27.35 & 2.61 \\
\textbf{2025-11-29.83$^\dagger$} & NIC & $J$, $H$, \Ks & 60 & 7, 8, 6 & 1.77 & 1.92 & \textbf{30.58} & 1.60 \\
\textbf{2025-12-02.85$^\dagger$} & SQUIDPOL & \Rc, \Ic & 300 & 7,7 & 1.85 & 1.89 & \textbf{30.54} & 1.48 \\
2025-12-03.86 & SQUIDPOL & \Ic & 300 & 8 & 1.87 & 1.88 & 30.44 & 1.40 \\
2025-12-08.83 & SQUIDPOL & \Rc, \Ic & 300 & 6,7 & 2.00 & 1.84 & 29.37 & 1.38 \\
2025-12-18.72 & NIC & $J$, $H$ & 120 & 3, 5 & 2.27 & 1.80 & 24.63 & 1.46 \\
2025-12-21.80 & SQUIDPOL & \Rc & 300 & 7 & 2.36 & 1.80 & 22.68 & 1.95 \\
2025-12-25.81 & SQUIDPOL & \Rc & 300 & 7 & 2.48 & 1.82 & 19.70 & 1.13 \\
2026-01-07.80 & HONIR & \Rc, \Ic & 120 & 3, 2 & 2.88 & 1.98 & 9.66 & 1.19 \\
2026-01-08.54 & MSI & \Rc & 60, 120 & 8 & 2.90 & 2.00 & 9.10 & 1.60 \\
\enddata

\tablenotetext{}{
\textbf{Note.—}
All quantities are evaluated at the midpoint of the observation period.
(1) UT date.
(2) Instrument used for imaging polarimetry.
(3) Filters used in each observation.
(4) Individual exposure time.
(5) Number of datasets used for each filter.
(6) Heliocentric distance.
(7) Geocentric distance.
(8) Phase angle.
(9) Airmass.
The ephemerides were obtained from the JPL Horizons system (\citealt{JPLHorizons2025}).
\textbf{Bolded rows} ($^\dagger$) indicate observations with both visible and NIR filters obtained at similar phase angles.
}
\end{deluxetable*}

\section{Results}
\begin{figure*}[htbp]
\centering
\includegraphics[width=\textwidth]{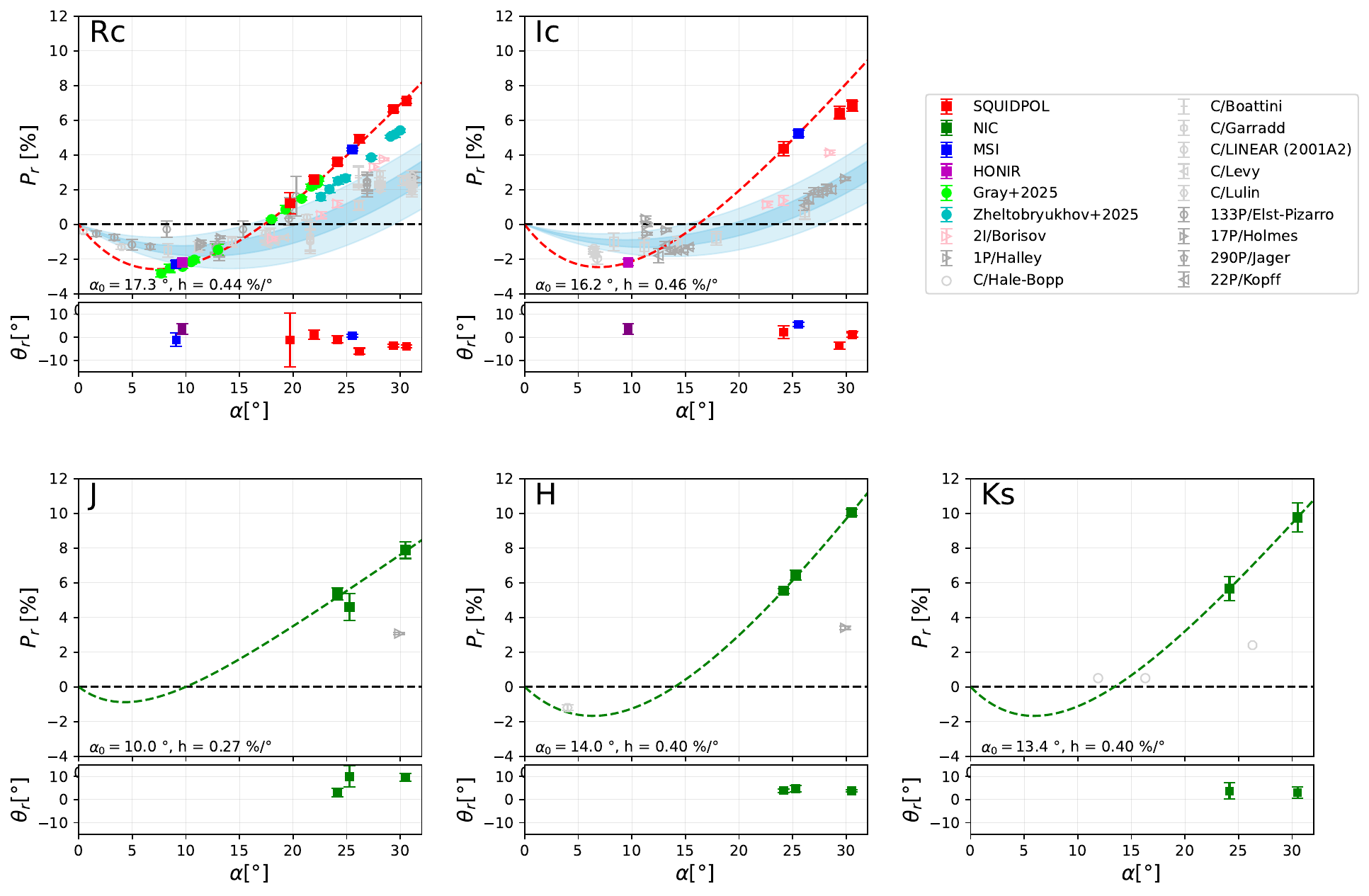}
\caption{
Polarization degree ($P_\mathrm{r}$) and polarization position angle ($\theta_\mathrm{r}$) of 3I/ATLAS as a function of phase angle, measured in five bands (\Rc, \Ic, $J$, $H$, and \Ks).
Colored symbols denote our visible and near-infrared imaging polarimetric data obtained with different instruments, while orange symbols show previously published measurements of 3I/ATLAS \citep{2025ApJ...992L..29G,2025RNAAS...9..338Z}.
Gray symbols represent polarimetric measurements of Solar System comets compiled in the Database of Comet Polarimetry (DBCP; \citealt{Kiselev2017}), and the pink triangles indicate the PPC of the interstellar comet 2I/Borisov \citep{2021NatCo..12.1797B}.
The dashed curves show the best-fit linear--exponential models (Equation~\ref{eq:pr}) to our data.}
\label{fig:PPC}
\end{figure*}

Figure~\ref{fig:PPC} shows the PPCs of 3I/ATLAS measured in the $R_\mathrm{C}$, $I_\mathrm{C}$, $J$-, $H$- and \Ks-bands.  The $R_\mathrm{C}$-band observations provide the highest sampling frequency, covering a phase-angle range of approximately $\alpha \simeq 9\degr$--$32\degr$ during post-perihelion orbital phase.
For comparison, previously published polarimetric data of 3I/ATLAS
(\citealt{2025ApJ...992L..29G}, obtained at heliocentric distances of $r_H \simeq 2.6$--4.0~au before crossing the water snow line,
and \citealt{2025RNAAS...9..338Z}, obtained at $r_H \simeq 1.4$--1.7~au after crossing the water snow line),
as well as PPCs of other SSCs, are also shown in the same figure. The latter observations were conducted at heliocentric distances where the ambient temperature is sufficiently higher than the sublimation temperature of water ice, corresponding qualitatively to the region interior to the water snow line.

To investigate the parameters describing the PPC, we utilized the empirical linear-exponential function (\citealt{Muinonen2002}; \citealt{Kaasalainen2003}), reparameterized as \citealt{2024A&A...684A..81B}: 
\begin{equation}\label{eq:pr}
  P_r(\alpha) 
    = h\frac{(1 - e^{-\alpha_0/k})\alpha - (1 - e^{-\alpha/k})\alpha_0}{1 - (1 + \alpha_0/k) e^{-\alpha_0/k}} ~,
\end{equation}
where $h := \left . \frac{dP_r}{d\alpha} \right |_{\alpha=\alpha_0} = P_r^\prime(\alpha=\alpha_0)$ is the so-called polarimetric slope $[\%/\degr]$ and $k \, [\degr]$ is a scale parameter. In this reparameterized form, the inversion angle $\alpha_0$ is obtained directly as a fit parameter.

As already reported in \citet{2025ApJ...992L..29G}, the PPC of 3I/ATLAS exhibits a deeper negative polarization branch (NPB) than any of the SSCs included for comparison. Our \Rc-band measurements connect smoothly with the PPC reported by \citet{2025ApJ...992L..29G}, while they show a clearly different PPC profile compared to the results of \citet{2025RNAAS...9..338Z}, which were obtained at a similar epoch (this difference is discussed in Section~\ref{s:disc}).

The \Ic-band PPC shows behavior broadly consistent with that in the \Rc-band. 
At near-infrared wavelengths ($J$, $H$, and \Ks-bands), the degree of polarization increases monotonically with phase angle over the observed range. However, due to the limited number of measurements at small phase angles ($\alpha \lesssim 23\degr$), it is difficult to determine an inversion angle at these wavelengths. As in the \Rc-band case, the depth of the NPB at all observed wavelengths is significantly deeper than that of typical SSCs, indicating that the polarimetric behavior of 3I/ATLAS is distinctive across all observed wavelengths.

Furthermore, the PPCs obtained during the inbound from \citep{2025ApJ...992L..29G} and outbound phases from this study are mutually consistent within the observational uncertainties in all bands, and no significant temporal variation in the polarimetric properties is detected over the observed period. These measurements were obtained at comparable heliocentric distances for a given phase angle, allowing a direct comparison between the two orbital phases.

\begin{figure*}[htbp]
\centering
\includegraphics[width=\textwidth]{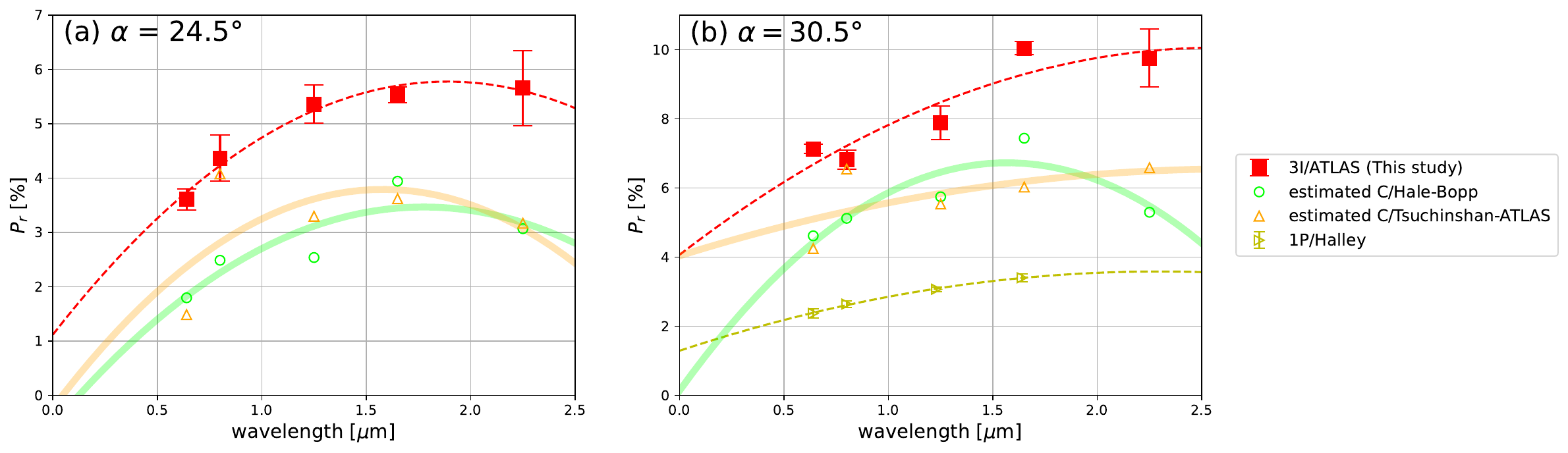}
\caption{Spectral dependence of the polarization degree of 3I/ATLAS at $\alpha = 24.2\degr$--$24.9\degr$ and $30.5\degr$, compared with those of comet C/Hale--Bopp, C/Tsuchinshan--ATLAS, and 1P/Halley.
For the comparison comets, polarization degrees in each band  are taken from the literature at similar phase angles and, when necessary, are estimated by extrapolation (Appendix~B; \citealt{Kikuchi1987,Kiselev2017,2025ApJ...983L..19L}).
The dashed and solid curves show Chebyshev polynomial fits.
}
\label{fig:PCC}
\end{figure*}

Figure~\ref{fig:PCC} shows the PCC of 3I/ATLAS at phase angles where polarimetric measurements are available in all observed bands ($\alpha = 23.4\degr$ and $30.5\degr$). Because the signal-to-noise ratio of the \Ks-band data obtained on 2025 December 18 is insufficient, the corresponding \Ks-band data point is not included in this figure.
At shorter wavelengths (0.6--1.2~\micron), $P_\mathrm{r}$ increases with wavelength. This wavelength dependence is consistently observed at both phase angles and represents a systematic trend exceeding the measurement uncertainties.
At longer wavelengths, the polarization does not continue to increase monotonically from the visible range. Instead, it reaches a maximum at wavelengths of approximately 1.5--2.0~\micron and remains nearly constant or shows a slight decrease beyond 1.7--2.2~\micron. 

\section{Discussion}\label{s:disc}

\subsection{Reliability of our data}
The polarization degrees measured in this study differ from those reported by \citet{2025RNAAS...9..338Z}, despite the observations being obtained at a similar epoch. This motivated a careful assessment of the reliability of our polarimetric measurements. We performed a comprehensive calibration program, including observations of polarized and unpolarized standard stars to determine the instrumental polarization, position-angle offset, and their uncertainties. The polarimetric efficiency was measured using a wire-grid filter, and the total error budget was evaluated by accounting for both statistical and systematic uncertainties (see Appendix~A). We also conducted a cross calibration between the newly deployed SQUIDPOL instrument and the Pirka/MSI system, which has a well-established polarimetric accuracy of $\sigma_P \simeq 0.1$--$0.2\,\mathrm{\%}$. The good agreement between these independent measurements demonstrates that our results are robust within the quoted uncertainties.

In addition, we considered the possible influence of gas depolarization in our data. Based on the spectroscopic measurements by \citet{2026arXiv260116983H}, the CN emission contributes approximately 1\% of the total \Rc-band flux around November 16. This small line-to-continuum ratio is broadly consistent with the spectrum obtained on 2025 August 15 by SPHEREx \citep{2025arXiv251207318L}, although the observing epochs are slightly different. Although CN emission may possess some intrinsic polarization \citep[e.g.,][]{2018A&A...620A.161K}, it is generally much lower than that of dust. Assuming that the CN emission is completely unpolarized, we estimate a conservative upper limit on the depolarization effect. In that case, the maximum depolarization would be only $\sim$0.04\% (i.e., $0.01 \times 4.32\%$), which is significantly smaller than our typical measurement uncertainty of $\sim$0.2\%. Although other emission species (e.g., NH$_2$) may also contribute in the \Rc-band, CN is generally brighter than NH$_2$ \citep{2018A&A...620A.161K}, suggesting that their contribution is even smaller. Therefore, gas contamination in the \Rc\- band can safely be considered negligible.

For the \Ic\-band, the red CN system may be relatively stronger at longer wavelengths \citep{2018A&A...620A.161K}. While a minor depolarization effect cannot be entirely excluded in the \Ic\-band, the polarization degrees in this band appear slightly lower than the expected phase-angle trend near perihelion (Figure 1). Nevertheless, for most of our observations, the overall gas contribution is expected to remain small compared to the measured polarization amplitude.

\subsection{Interpretation of PPC}

The PPCs presented here place important constraints on the origin of the polarimetric properties of 3I/ATLAS. The continuous connection of the PPCs measured inside and outside the water snow line, together with the agreement between the inbound and outbound PPCs across all observed bands, indicates that the polarimetric behavior of 3I/ATLAS remained stable over the observed orbital arc.


The unusually large polarization amplitudes of 3I/ATLAS therefore point to dust properties that differ from those of typical SSCs. Assuming dust ejection velocities of 1--100~m~s$^{-1}$ for 10--1000~\micron-sized grains typical of Solar System cometary dust \citep{2007Icar..189..169I}, the dust sampled within the photopolarimetric aperture (comet-centric radius of $\rho = 15{,}000-20{,}000\,\mathrm{km}$) would have been released $\sim$2--200~days prior to the observations.
At a heliocentric distance of $\sim$2~au, a typical dirty water-ice dust particle (grain size $\sim$10--1000~\micron) would have completely sublimated within this timescale if it were composed of water ice \citep{1986A&A...164..397M}. This implies that the dust observed at $\alpha \gtrsim 25\degr$ is largely refractory.
The consistency of the PPCs between the inbound and outbound orbital phases further suggests that dust with nearly uniform polarimetric properties was continuously released throughout the observed period.

\subsection{Interpretation of PCC}


The non-monotonic wavelength dependence of polarization observed for 3I/ATLAS provides important constraints on the characteristic size scale of its scattering particles. In particular, the presence of a polarization maximum at near-infrared wavelengths ($\sim$1.5--2.0~\micron) indicates that the dominant scattering units are not single compact grains, but aggregates composed of smaller constituents.

Similar behavior has been reported for several well-studied Solar System comets, including 1P/Halley, C/West, C/Hale--Bopp, and C/2023~A3 (Tsuschinshan--ATLAS), observed over a wide range of phase angles and heliocentric distances. In these cases, the PCC deviates from a simple monotonic increase with wavelength and instead exhibits a maximum in the near-infrared \citep{2025ApJ...983L..19L}. The recurrence of this feature across different comets suggests that such a polarization maximum is a common signature of light scattering by dust aggregates composed of submicron-sized monomers, consistent with theoretical and laboratory studies (e.g., \citealt{2025ApJ...983L..19L}).

\subsection{Implications for the Nature of Dust in 3I/ATLAS}

Taken together, the distinctive PPC of 3I/ATLAS is most naturally explained by the intrinsic optical properties of refractory dust particles, rather than by the presence or absence of volatile materials. Although a small number of SSCs (e.g., C/1995~O1 [Hale--Bopp], 2P/Encke, and 252P/LINEAR) exhibit unusually high polarization degrees, these cases are generally attributed to special circumstances such as large nucleus size or advanced evolutionary processing. Such scenarios are unlikely to apply to 3I/ATLAS, which has a much smaller nucleus ($D \approx 1$~km; \citealt{2025ApJ...990L...2J}) and is presumably experiencing stellar heating for the first time.


A particularly intriguing result is that the characteristic monomer size inferred from the PPC and PCC of 3I/ATLAS, based on comparisons with previous theoretical and observational studies of cometary dust polarization, is comparable to that of dust aggregates in SSCs ($\sim$0.2--0.5~\micron, \citealt{2004come.book..577K}) and to typical grain sizes inferred for interstellar dust from extinction studies \citep{2003ARA&A..41..241D}. This suggests that, while the refractory dust composition in the planet-forming environment of 3I/ATLAS may differ from that of the Solar System, the fundamental size scale of the basic scattering units is likely universal within our Galaxy.


\section{Summary}
We present the post-perihelion polarimetric data of 3I/ATLAS, covering a phase angle range of $\alpha \simeq 9\degr$--$32\degr$ in both optical ($R_\mathrm{C}$ and $I_\mathrm{C}$) and near-infrared ($J$-, $H$-, and \Ks-bands) wavelengths. Our optical polarimetric results were cross-calibrated between the SQUIDPOL instrument and the Pirka/MSI system. The key findings are as follows: (i) the PPC exhibits a large polarization amplitude and shows good consistency with the pre-perihelion data \citep{2025ApJ...992L..29G}, and (ii) the PCC displays behavior similar to Solar System comets. These results suggest that 3I/ATLAS consists of refractory grains with an aggregate structure, similar in size to typical Solar System comets. Combining the PPC and PCC results, we interpret these findings as indicating that, although compositional differences may exist in refractory dust, the fundamental size of the scattering units may be broadly similar to that inferred for typical Solar System comets.

\section*{Data Availability}
The availability of the observational data listed in Table 1 is as follows. (1) SQUIDPOL data: The polarimetric data products are publicly available via Zenodo\footnote{\url{https://zenodo.org/records/18627153}} with a DOI (\url{10.5281/zenodo.18627153}). The archive includes the calibration data, observational data of 3I/ATLAS, and associated metadata necessary for reuse. (2) Pirka Telescope data: The Pirka observations are not stored in a public archive. The observational data products are available upon reasonable request to the co-authors: Hisayuki Kubota (\url{hkubota@sci.hokudai.ac.jp}) and Seiko Takagi (\url{seiko@ep.sci.hokudai.ac.jp}). (3) Nishiharima Astronomical Observatory and Higashi-Hiroshima Observatory data: These observational data will be made publicly available through the SMOKA (Subaru–Mitaka–Okayama–Kiso Archive System) archive after the standard 18-month proprietary period. Readers are encouraged to access the data via SMOKA once they are released \citep{2002ASPC..281..298B}.

\begin{acknowledgments}
This research was supported by a grant from the Korean National Research Foundation (NRF) (MEST), funded by the Korean government (No. 2023R1A2C1006180).
We thank the anonymous referee for constructive comments that helped improve the clarity of the manuscript.
\end{acknowledgments}

\bibliography{choi2026_rev1}{}
\bibliographystyle{aasjournalv7}

\appendix

\section{Appendix: Data Reduction and Error Analysis}

\begin{deluxetable*}{cccccc}[h]
\setcounter{table}{0}
\renewcommand{\thetable}{A\arabic{table}}
\label{tab:polcoeff}
\tabletypesize{\small}
\tablewidth{\textwidth}
\tablecolumns{6}
\setlength{\tabcolsep}{6pt}
\tablecaption{Instrumental polarization calibration coefficients.}
\tablehead{
\colhead{Instrument} & \colhead{Filter} & \colhead{$q_\text{inst}$ [\%]} &
\colhead{$u_\text{inst}$ [\%]} & \colhead{$p_\text{eff}$ [\%]} &
\colhead{$\theta_\text{off}$ [$\degr$]}
}
\startdata
NIC & $J$ & -0.00 $\pm$ 0.29 & -0.01 $\pm$ 0.29 & 98 $\pm$ 6 & 0.5 $\pm$ 1.3 \\
 & $H$  & 0.03 $\pm$ 0.52 & -0.03 $\pm$ 0.55 & 95 $\pm$ 7 & 1.3 $\pm$ 3.1 \\
 & \Ks  & -0.02 $\pm$ 0.30 & -0.07 $\pm$ 0.31 & 92 $\pm$ 12 & -0.7 $\pm$ 6.3 \\\hline
MSI & \Rc & 0.645 $\pm$ 0.275 & 0.407 $\pm$ 0.094 & 99.09 $\pm$ 0.02 & 2.32 $\pm$ 0.43 \\
 & \Ic & 0.098 $\pm$ 0.140 & -0.126 $\pm$ 0.161 & 99.09 $\pm$ 0.02 & 1.43 $\pm$ 0.32 \\\hline
SQUIDPOL-A & \Rc & -0.26 $\pm$ 0.32 & -0.38 $\pm$ 0.31 & 96.15 $\pm$ 0.48 & 10.27 $\pm$ 0.14 \\
 & \Ic & -1.75 $\pm$ 0.55 & -0.68 $\pm$ 0.30 & 94.77 $\pm$ 0.69 & 10.45 $\pm$ 0.12 \\\hline
SQUIDPOL-B & \Rc & 0.04 $\pm$ 0.11 & -0.01 $\pm$ 0.14 & 97.35 $\pm$ 0.03 & 10.27 $\pm$ 0.14 \\
 & \Ic & -0.04 $\pm$ 0.18 & 0.09 $\pm$ 0.08 & 96.70 $\pm$ 0.09 & 10.45 $\pm$ 0.12 \\\hline
\enddata
\end{deluxetable*}

\begin{deluxetable*}{ccccccccccc}[h]
\renewcommand{\thetable}{A\arabic{table}}
\label{tab:sumpol}
\addcontentsline{toc}{section}{Appendix}
\tabletypesize{\small}
\tablewidth{\textwidth}
\setlength{\tablecolumns}{9}
\setlength{\tabcolsep}{6pt}
\tablecaption{Summary of aperture polarimetry.}
\tablehead{
\colhead{UT Date} & \colhead{$\alpha$[\degr]} & \colhead{$r_H$[au]} & \colhead{Instrument} & \colhead{Filter} & \colhead{$P$ [\%]} & \colhead{$\sigma_P$ [\%]} & \colhead{$\theta$ [\degr]} & \colhead{$\sigma_\theta$ [\degr]} & \colhead{$P_\mathrm{r}$ [\%]}& \colhead{$\theta_\mathrm{r}$ [\degr]} \\
\colhead{} & \colhead{} & \colhead{} & \colhead{} & \colhead{(1)} & \colhead{(2)} & \colhead{(3)} & \colhead{(4)} & \colhead{(5)} & \colhead{(6)}
}
\startdata
2025-11-13.85 & 24.16 & 1.47 & NIC & $J$ & 5.39 & 0.35 & 68.11 & 1.88 & 5.36 & 3.05 \\
 & & & & $H$ & 5.58 & 0.14 & 68.95 & 0.73 & 5.53 & 3.90 \\
 & & & & \Ks & 5.70 & 0.69 & 68.78 & 3.47 & 5.65 & 3.72 
\\ \hline
2025-11-14.84 & 24.87 & 1.48 & SQUIDPOL & \Rc & 3.61 & 0.19 & 23.83 & 1.53 & 3.61 & -0.99 \\
 & & &  & \Ic & 4.40 & 0.42 & 27.05 & 2.73 & 4.37 & 2.24 
\\\hline
2025-11-15.85 & 25.55 & 1.50 & MSI & \Rc & 4.32 & 0.08 & 25.27 & 0.54 & 4.32 & 0.67 \\
 & & &  & \Ic & 5.33 & 0.19 & 30.16 & 1.02 & 5.23 & 5.57 \\\hline
2025-11-18.85 & 27.35 & 1.55 & SQUIDPOL & \Rc & 5.05 & 0.24 & 18.43 & 1.35 & 4.93 & -6.06 \\\hline
2025-11-29.83 & 30.58 & 1.77 & NIC & $J$ & 8.35 & 0.48 & 75.91 & 1.66 & 7.88 & 9.68 \\
 & && & $H$ & 10.13 & 0.19 & 70.01 & 0.54 & 10.04 & 3.78 \\
 & && & \Ks & 9.81 & 0.84 & 68.14 & 2.44 & 9.76 & 3.09 \\ \hline
2025-12-02.85 & 30.54 & 1.85 & SQUIDPOL & \Rc & 7.20 & 0.14 & 19.71 & 0.55 & 7.13 & -3.93 \\
 & && & \Ic & 6.83 & 0.28 & 24.81 & 1.17 & 6.82 & 1.17 \\\hline
2025-12-03.86 & 30.44 & 1.87 & SQUIDPOL & \Ic & 6.93 & 0.24 & 24.61 & 0.99 & 6.92 & 0.98 \\\hline
2025-12-08.83 & 29.37 & 2.00 & SQUIDPOL & \Rc & 6.71 & 0.16 & 19.65 & 0.70 & 6.65 & -3.60 \\
 & && & \Ic & 6.51 & 0.35 & 19.61 & 1.54 & 6.45 & -3.65 \\\hline
2025-12-18.72 & 24.63 & 2.27 & NIC & $J$ & 4.89 & 0.78 & 77.45 & 4.56 & 4.60 & 9.98 \\
 & && & $H$ & 6.52 & 0.30 & 72.14 & 1.31 & 6.43 & 4.67 \\\hline
2025-12-21.80 & 22.68 & 2.36 & SQUIDPOL & \Rc & 2.60 & 0.18 & 22.98 & 2.03 & 2.59 & 1.01 \\\hline
2025-12-25.81 & 19.70 & 2.48 & SQUIDPOL& \Rc & 1.34 & 0.55 & 20.41 & 11.75 & 1.23 & -1.19 \\\hline
2026-01-07.80 & 9.66 & 2.88 & HONIR & \Rc & 2.21&0.17&-12.94&2.25&-2.19&86.43\\
 & & &  & \Ic & 2.24&0.14&-16.23&1.78&-2.18&83.14\\\hline
2026-01-08.54 & 9.10 & 2.90 & MSI & \Rc & 2.31 & 0.23 & 111.83 & 2.90 & -2.30 & 91.13 
\enddata

\tablenotetext{}{NOTE--(1) Observed polarization degree. (2) Standard deviation of $P$. (3) Position angle of the strongest electric field vector. (4) Standard deviation of $\theta$. (5) Polarization degree relative to the scattering plane. (6) Position angle of the strongest electric field vector relative to the scattering plane normal.}
\end{deluxetable*}

The preprocessing includes bias subtraction, dark subtraction, and flat-field correction, together with cosmic-ray rejection. 
For NIC data, the vertical pattern noise was corrected at the raw-image level using the NIC data reduction package NICpolpy \citep{2022S&G....5....4B}. 
After preprocessing, fluxes were extracted using aperture photometry. In the case of SQUIDPOL, the large seeing and tracking errors broadened the point-spread function (PSF) of 3I/ATLAS. Therefore, we adopted an aperture radius of approximately 30 pixels to collect as much signal from 3I/ATLAS as possible, which physically corresponds to about $1.5\times10^4-2.0\times10^4$ km at the location of the comet and thus includes light not only from the nucleus but also from the surrounding coma. This aperture size was confirmed with the analysis based on aperture size dependency on Figure \ref{fig:A1}.
\renewcommand{\thefigure}{A\arabic{figure}}
\setcounter{figure}{0}

\begin{figure*}[htbp]
\centering
\includegraphics[width=\textwidth]{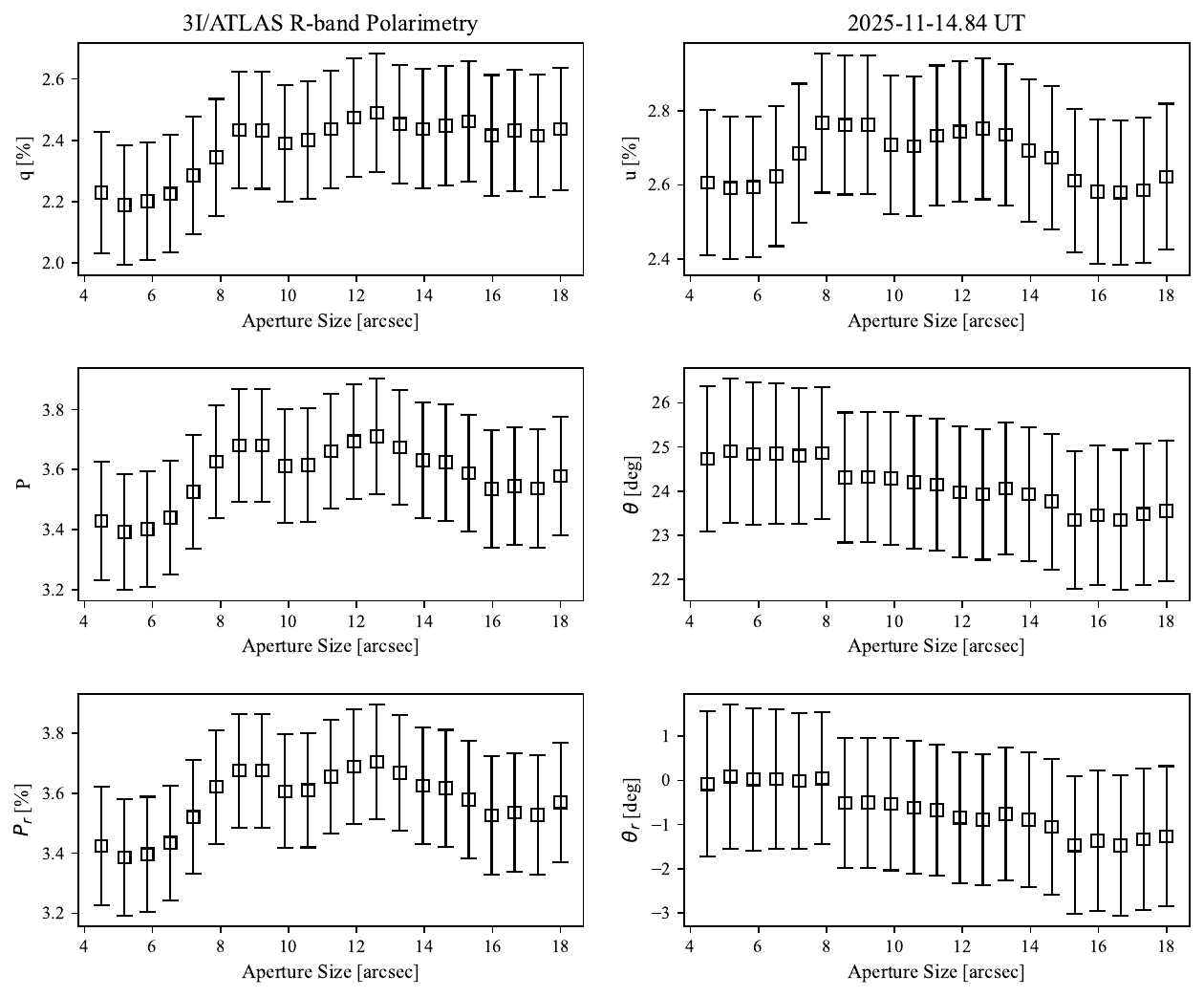}
\caption{Aperture size dependency of polarimetric analysis for observed data on Nov.14}
\label{fig:A1}
\end{figure*}

The linear polarization degree $P$ and polarization position angle $\theta$ are defined as

\begin{equation}
P = \sqrt{(Q/I)^2+(U/I)^2}
\tag{A1}\label{A1}~~,
\end{equation}

\begin{equation}
\theta = \frac{1}{2}\arctan\left(\frac{U}{Q}\right).
\tag{A2}\label{A2}
\end{equation}

The observed Stokes parameters were calibrated for instrumental polarization ($q_\mathrm{int}$, $u_\mathrm{int}$), polarization efficiency ($p_\mathrm{eff}$), and position-angle offset ($\theta_\mathrm{offset}$) using the coefficients listed in Table~A\ref{tab:polcoeff}. 
For SQUIDPOL, the calibration parameters were updated after a system upgrade in late December 2025. Accordingly, data obtained before 2025 December 25 were calibrated using the SQUIDPOL-A coefficients, while data obtained on and after that date were calibrated using SQUIDPOL-B.
After calibration, the linear polarization degree $P$ and the electric vector position angle $\theta$ were calculated from the extracted Stokes parameters $Q$, $U$, and $I$ using Equations~(\ref{A1}) and (\ref{A2}).

To correct the inherent bias of the observed linear polarization degree due to random noise, we adopted the debiased polarization degree using Equation~(\ref{A3})(\citealt{1974ApJ...194..249W}).
\begin{equation}
P_{\text{debias}} = \sqrt{P^2-\sigma_P^2}.
\tag{A3}\label{A3}
\end{equation}

The polarization degree relative to the scattering plane, $P_\mathrm{r}$, and the corresponding position angle, $\theta_\mathrm{r}$, were derived as

\begin{equation}
P_\mathrm{r} = P_{\text{debias}}\cos{2\theta_r},
\tag{A4}\label{A4}
\end{equation}

\begin{equation}
\theta_\mathrm{r} = \theta - (\phi\pm90\degr),
\tag{A5}\label{A5}
\end{equation}

\noindent where $\phi$ is the position angle of the scattering plane at the time of observation for each instrument. 
We selected the sign of $\phi$ which satisfies $-90 \degr < \theta_r < +90 \degr$.

Table~A\ref{tab:sumpol} summarizes the results of aperture polarimetry obtained with SQUIDPOL, MSI (\Rc\ and \Ic), and NIC ($J$, $H$, and \Ks).

\section{PCC Estimation for C/Hale--Bopp and C/Tsuchinshan--ATLAS}

In Figure~\ref{fig:PCC}, the data points for two comets, C/Hale--Bopp and C/Tsuchinshan--ATLAS, are extrapolated values, because no polarimetric measurements are available at phase angles of $\alpha \sim 24\degr$ and $\alpha \sim 30\degr$ for these comets.
Using Equation~(\ref{eq:pr}), we constructed polarization phase curves (PPCs) in each filter (\Rc, \Ic, $J$, $H$, and \Ks) and extrapolated the corresponding polarization degrees $P_\mathrm{r}$ at the required phase angles.

Since Equation~(\ref{eq:pr}) reliably reproduces observed polarization behavior only at small phase angles, the fitting was performed using data points with phase angles smaller than $70\degr$.
Figures~\ref{fig:B1} and \ref{fig:B2} show the resulting PPC fits, where the red and blue symbols indicate the extrapolated polarization degrees at $\alpha = 24.5\degr$ and $\alpha = 30.5\degr$, respectively.

\renewcommand{\thefigure}{B\arabic{figure}}
\setcounter{figure}{0}

\begin{figure*}[htbp]
\centering
\includegraphics[width=0.85\textwidth]{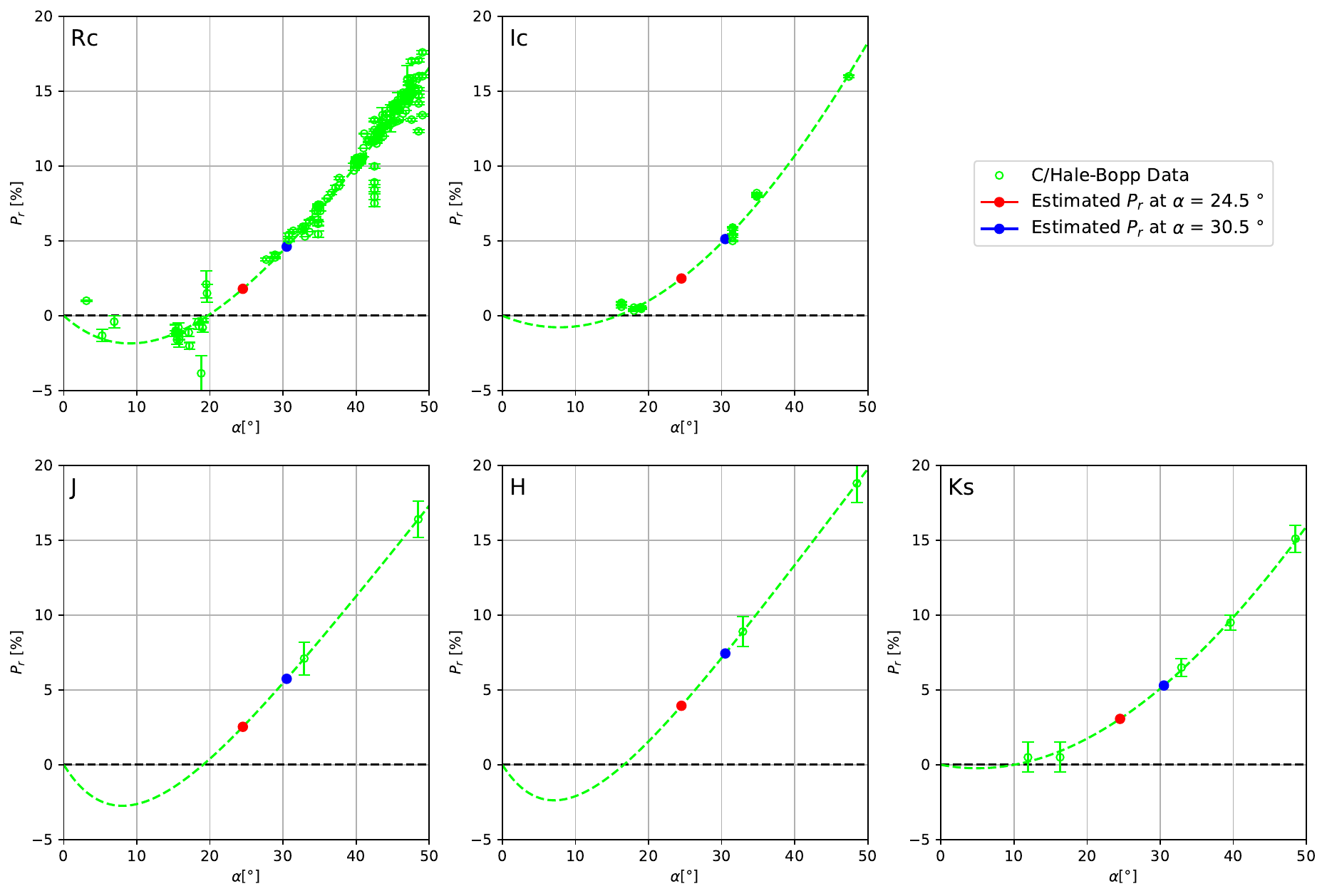}
\caption{PPC of C/Hale-Bopp within each filter}
\label{fig:B1}
\end{figure*}

\begin{figure*}[htbp]
\centering
\includegraphics[width=0.85\textwidth]{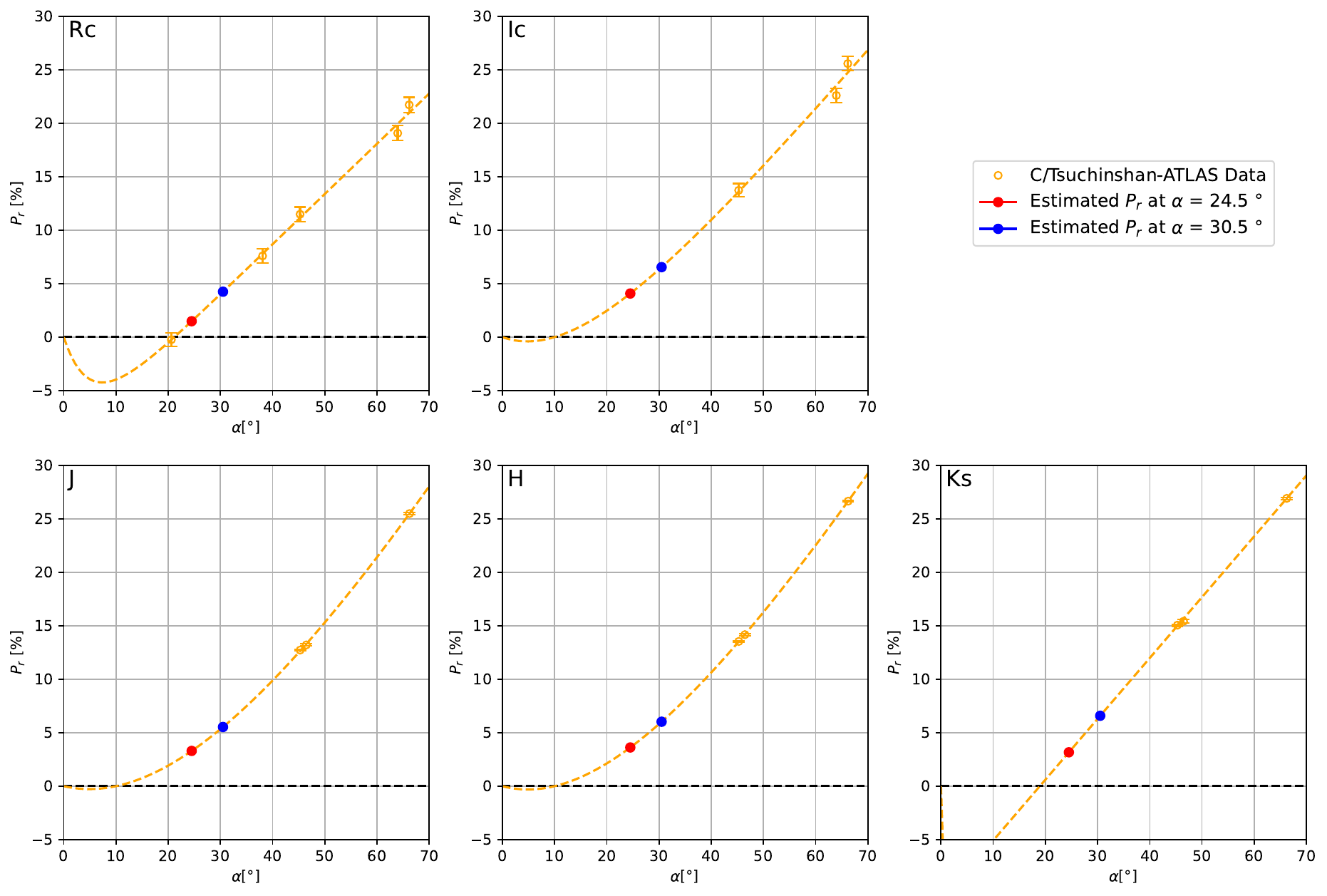}
\caption{PPC of C/Tsuchinshan-ATLAS within each filter}
\label{fig:B2}
\end{figure*}

\end{document}